\documentclass[prl,twocolumn,groupedaddress]{revtex4-1}
\usepackage{graphicx}

\begin{document}
				\title{
Transition and formation of the torque pattern of undulatory locomotion in resistive force dominated media
				}
							
		\author{
Tingyu Ming}

\affiliation{
		Beijing Computational Science Research Center \\
		Haidian District, Beijing 100193, China }
\author{
Yang Ding}\email{dingyang@csrc.ac.cn}
\affiliation{
		Beijing Computational Science Research Center \\
		Haidian District, Beijing 100193, China }


				\date{\today} 

\begin{abstract}
In undulatory locomotion, torques along the body are required to overcome external forces from the environment and bend the body. These torques are usually generated by muscles in animals and closely related to muscle activations. In previous studies, researchers observed a single traveling wave pattern of the torque or muscle activation, but the formation of the torque pattern is still not well understood. To elucidate the formation of the torque pattern required by external resistive forces and the transition as kinematic parameters vary, we use simplistic resistive force theory models of self-propelled, steady undulatory locomotors and examine the spatio-temporal variation of the internal torque. We find that the internal torque has a traveling wave pattern with a decreasing speed normalized by the curvature speed as the wave number (the number of wavelengths on the locomotor's body) increases from 0.5 to 1.8. As the wave number increases to 2 and greater values, the torque transitions into a two-wave-like pattern and complex patterns. Using phasor diagram analysis, we reveal that the formation and transitions of the pattern are consequences of the integration and cancellation of force phasors.
\end{abstract}

\maketitle

\section{Introduction}
Undulatory locomotion is a common way for animals to move in various environments (e.g. spermatozoa in water \cite{gray1955propulsion}, sandfish in sand~\cite{maladen2009undulatory}, snakes on land~\cite{hu2009mechanics,guo2008limbless}, and fish in water~\cite{wardle1995tuning}, for reviews, see \cite{lauga2009hydrodynamics,cohen2010swimming,alexanderbook} ) and a popular mode of locomotion for bio-inspired robots~\cite{tesch2009parameterized,crespi2008online,maladen2011undulatory}. This type of locomotion consists of bending the body or some portion of the body to form a traveling wave in the direction opposite to the motion direction to generate propulsion. For organisms in environments dominated by resistive forces, such as spermatozoa swimming at low Reynolds numbers (Re) and snakes slithering on the ground, how the propulsive forces from the environments are generated is quite well understood~\cite{cohen2010swimming,goldman2010wiggling}.

To bend the body and generate propulsion, internal torques (bending moments) are required to overcome both the restoring forces and damping forces of the body and the external forces from the resistance of surrounding media. For macroscopic animals such as eels and snakes, the internal torques are generated by muscle forces acting on the body. Therefore, in previous theoretical and computational studies, spatio-temporal torque patterns were used to explain and predict muscle activation patterns~\cite{cheng1994bending,cheng1998continuous,hess1984fast}. Ongoing interdisciplinary research over the past several decades has provided a general overview of the torque and muscle activation: they both exhibit traveling wave patterns from head to tail. However, the waves of the muscle activation and the torque travel faster than the wave of the curvature, which is a phenomenon known as  neuromechanical phase lags~\cite{wardle1995tuning,butler2015consistent}. Consequently, muscles activate after they begin to shorten in the anterior part of the body, and muscles begin to activate before they begin to shorten in the posterior part of the body. 

By imposing kinematics and considering contributions from the resistance of the environment and the passive body properties of fish (e.g. saithe and lamprey), qualitative agreements between predicted torque patterns and muscle activation patterns have been achieved~\cite{cheng1994bending,cheng1998continuous,hess1984fast}. For the relatively simple case of the sandfish lizard swimming in sand, where resistive forces dominate and the body is nearly uniform, a quantitative agreement has been obtained using resistive force theory (RFT)~\cite{ding2013emergence}. However, how the torque pattern is formed and whether the pattern is always a traveling wave are still open questions. 
 \begin{table}
\caption{\label{tab1}Wave numbers observed in nature}
  \resizebox{0.5 \textwidth}{!}{%
  \begin{tabular}{| c | c | c | c |c |c|c|}
     \hline
     Organism & Spermatozoon  & Nematode & Snake  &  Eel & Scup & Sandfish \\  \hline
     Wave number \footnote{When only amplitude and wavelength are given in the reference, we assume that the motion is sinusoidal and approximated the wave number as $\xi=L/\int_0^\lambda{\sqrt{1+B^2\sin^2{x}}\mathrm{d}x}$, where $L$ is the body length, $B$ is the undulation amplitude, and $\lambda$ is the wavelength.}& 1.25-1.4 &0.55-1.31 & 1.60 (on ground), 3.5 (in sand) & 1.7 & 0.65 & 1.0 \\ \hline
	Source & \cite{gray1955propulsion,brokaw1965non} & \cite{fang2010biomechanical,berri2009forward} & \cite{hu2009mechanics,sharpe2015locomotor} & \cite{wardle1995tuning} & \cite{wardle1995tuning} & \cite{maladen2009undulatory} \\ \hline
  \end{tabular}
   }
\end{table}
   
Another approach for studying the mechanics of undulatory locomotion is to start with the internal forces/torques and observe the kinematics as a result of the couplings between the internal drives, passive body properties and external environments. Two closely related kinematic parameters are the wavelength and the wave number (the number of wavelengths on the locomotor's body), which, in reality, vary in different species and for the same species in different environments (Tab.~\ref{tab1}). For example, the wavelength decreases as the viscosity increases in nematodes and spermatozoa~\cite{fang2010biomechanical,brokaw1966effects}. By imposing a neuron activation pattern, muscle forces, or a relationship between internal shear force and curvature, previous studies showed the trend of decreasing wavelength in spermatozoa and fish swimming when the relative strength of the external resistance to the internal driving forces/torques is reduced~\cite{johnson1979flagellar,tytell2010interactions,mcmillen2008nonlinear}. However, how the variations in internal torque, kinematics and other components interact with each other is still not well understood.

In robots using undulatory gaits, torques are generally generated directly by motors (e.g., \cite{choset2000design}), although new actuation mechanisms are emerging~\cite{yan2012novel,chu2012review,nguyen2009c}. Torque is also a convenient way for detecting unexpected forces and avoiding damage to robots~\cite{de2005sensorless}. A deep understanding of the features of torques such as their magnitudes, power output, and phase relationships with curvature in various configurations and environments is useful for designing driving systems~\cite{liljeback2012review,wright2007design}.

Here, we consider steady forward undulatory locomotion in resistive-forces-dominated media with simple kinematics and body shape. We show the basic torque pattern and its transitions to new patterns as the wave number increases. Further analysis reveals the formation of the torque pattern and the underlying mechanism of the transitions. 

\section{Model}
\begin{figure}[h]
	\begin{center}
    \includegraphics[width=0.45\textwidth]{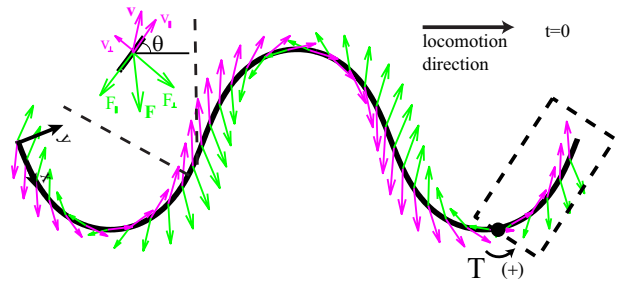}
  \caption{Diagram of the model. The black curve represents the body, the magenta arrows represent velocities, and the green arrows represent forces from the medium. The black dot is a representative point on the body at which the internal torque ($T$) is calculated from the forces in the dashed box. The sign and arrow indicate the direction of the torque. $\xi=1.5$, $A=7.54$, and $t=0$.}\label{fig:model}
  	\end{center}
\end{figure}

We consider an undulatory locomotor bending its slim and uniform body as a traveling serpentine wave in a plane (Fig.~\ref{fig:model}). We use body length as unit length and one undulation period $T_p$ as unit time. The curvature is prescribed as $\kappa=A \xi \sin[2\pi (\xi s+ t)]$, where $s \in [0\,\, 1]$ is the arc length measured from the tail, $A$ controls the undulation amplitude relative to the wavelength, $\xi$ is the wave number, and $t$ is the time. The speed of the curvature wave becomes $v_\kappa=1/\xi$. $A$ is set to $7.54$, which gives an amplitude-to-wavelength ratio ($\approx$0.24) that is close to experimentally observed ratios~\cite{hu2009mechanics,maladen2009undulatory,gray1955propulsion}. For every time instant, we use a body frame in which the tail end is at the origin and pointing toward the $x-$ axis. The tangent angle of a segment at $s$ to the $x+$ axis can be computed by integrating the curvature along the body: $\theta(s) = \int_0^s \kappa \mathrm{d}l$. The position of the segment can be computed as $\mathbf{r}(s)=(x,y)=( \int_0^s \cos(\theta) \mathrm{d}l,\int_0^s \sin(\theta) \mathrm{d}l)$. By taking the time derivative, the velocity $\mathbf{v}_b$ of a segment relative to the tail end can be computed. Assuming that the tail end is moving at velocity $\mathbf{v}_\mathrm{tail}$ and rotating at angular velocity $\omega$, the velocity at the body position $s$ in the lab frame becomes $\mathbf{v}=\mathbf{v}_b+\mathbf{v}_\mathrm{tail}+\omega \mathbf{e}_z \times \mathbf{r}$. 

To determine the motion of the body and the distribution of the force on the body, we use an RFT model similar to those in \cite{ding2013emergence,hu2009mechanics}. In the RFT model, the body is divided into infinitesimal segments. Assuming that the force ($\mathbf{F}(s)$) experienced by one segment is independent of other segments, the force can be calculated based on the geometry, orientation, and velocity of the segment. The perpendicular and parallel components of the force on a segment can be written as $F_\perp(v_\perp,v_\parallel)$ and $F_\parallel(v_\perp,v_\parallel)$, respectively, where $v_\perp$ and $v_\parallel$ are the perpendicular and parallel components of the segment velocity $\mathbf{v}$. We first consider the simplest case, in which the head drag is negligible and the forces are from viscous drag: $F_\perp=C_\perp v_\perp$ and $F_\parallel=C_\parallel v_\parallel$, where $C_\perp=2$ and $C_\parallel=1$ are the drag coefficients for a very thin cylinder~\cite{rodenborn2013propulsion}. The total external torque on the body can be computed as $\mathbf{F}_\mathrm{total}=\int_0^1 \mathbf{F}(l) \mathrm{d}l$ and $T_\mathrm{total}=\mathbf{e}_z\int_0^1 \mathbf{r}(l) \times \mathbf{F}(l) \mathrm{d}l$. We assume that inertia is negligible, which is a good approximation for micro-swimmers in fluids, crawlers on land, and swimmers in granular materials~\cite{gray1955propulsion,hu2009mechanics,maladen2009undulatory}. Under this assumption, the resultant net force $\mathbf{F}_\mathrm{total}$ and the net torque $T_\mathrm{total}$ related to the tail (reference) frame are both zero, from which $\mathbf{v}_\mathrm{tail}$ and $\omega$ can be determined. The motion and force distributions on the body are shown in Supplementary Video S1, and the MATLAB scripts of the computation are provided in the Supplementary Materials.


To compute the internal torque ($T$) required to overcome the external forces at a point on the body, we analyze the torque balance on the anterior side of the body at that point and simply find that $T(s)=-T^e(s)=-\int_{s}^1 [\mathbf{r}(l)- \mathbf{r}(s)]\times \mathbf{F}(l)\mathrm{d}l$ (see Fig.~\ref{fig:model} for an example), where $T^e$ is the total external torque from the anterior side of the body. Integrating over the posterior side of the body gives the same results. To compute the wave speed of the torque, we use the fitting function $\sqrt{2}\langle T(s) \rangle \sin[2\pi(s/\lambda_T+t)+\phi_T]$, where $\langle T(s) \rangle$ is the standard deviation of the torque at $s$, $\lambda_T$ is the wavelength of the torque wave along the body, and $\phi_T$ is a fitting parameter for the phase. The $\langle T(s) \rangle$ term is used to capture the variation in torque amplitude and the prefactor $\sqrt{2}$ comes from the ratio between the maximum and the standard deviation of the sine function. The fitting parameters $\lambda_T$ and $\phi_T$ are obtained from the best fitting of the torque. The speed of the torque wave is defined as $v_T=\lambda_T/T_p=\lambda_T$ and the speed ratio of the torque wave to the curvature wave is $v_T/v_\kappa=\lambda_T \xi$. 


\begin{figure*}
	\begin{center}
    \includegraphics[width=0.9\textwidth]{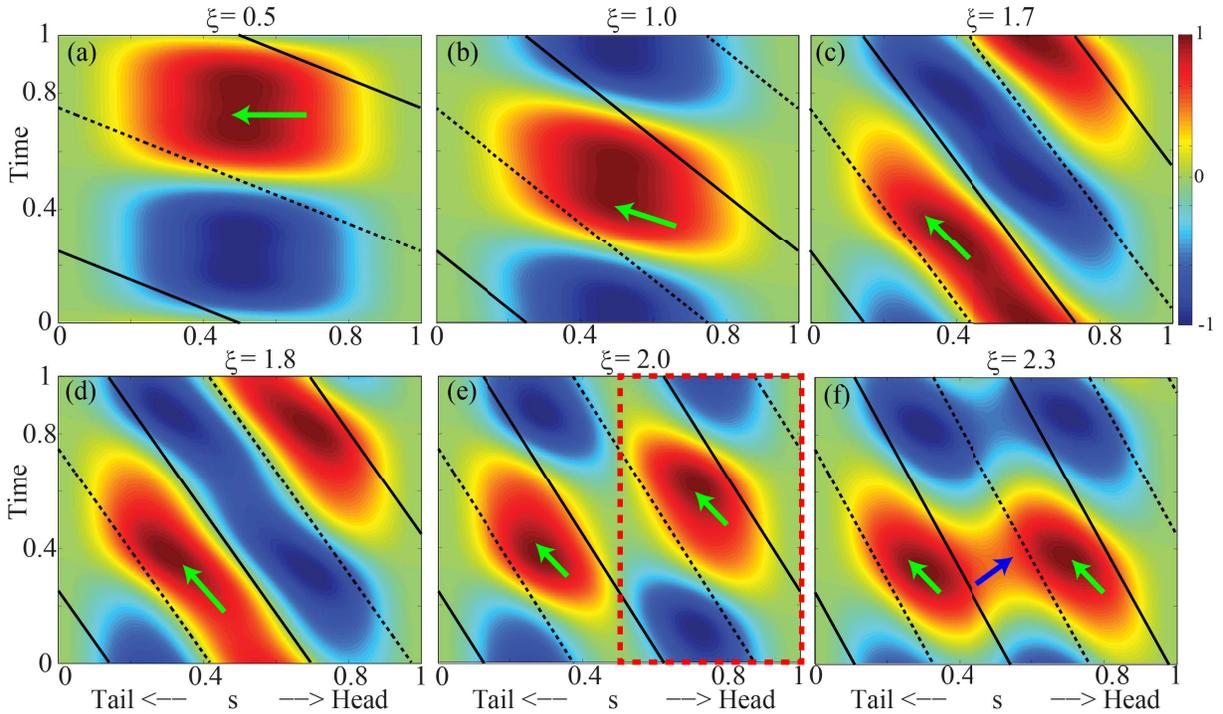}
  \caption{The internal torque as a function of body position and time for different wave numbers. The magnitude of the torque is normalized by the maximum torque and represented by color. The solid and dashed lines indicate the maximum and minimum curvatures, respectively. The red dotted box in (e) indicates the torque pattern that appears similar to the pattern in (b). The green and blue arrows indicate the local waves traveling posteriorly and anteriorly, respectively. $A=7.54$}\label{fig:torque}
  	\end{center}
\end{figure*}

\section{Results}

To focus on the torque pattern, we normalize the torque by its maximum value for each wave number $\xi$. We find that the torque exhibits a traveling wave pattern for $\xi<1.8$ (Fig.~\ref{fig:torque}a-d). In this regime, the amplitude of the torque is smaller near the ends and greater in the middle. As in previous studies, the torque wave travels faster than the curvature wave, and different phase lags between the curvature and the torque along the body are observed. When $\xi$ approaches 2, the magnitude of the torque in the middle of the body suddenly decreases, and a pattern of two apparently separated traveling waves forms (Fig.~\ref{fig:torque}e); each wave is similar to a wave with $\xi= 1$ and takes half of the body. For $\xi> 2$, the two waves merge as the wave direction near the middle of the body becomes the opposite direction of the curvature wave. The torque pattern is no longer one or two traveling waves (Fig.~\ref{fig:torque}f). Up to at least $\xi=11$, similar transitions occur by adding one additional traveling wave pattern near the middle of the body when $\xi$ reaches integer numbers. See Supplementary Video S2 for torque patterns with smaller $\xi$ increments.

\begin{figure}
	\begin{center}
    \includegraphics[width=0.48\textwidth]{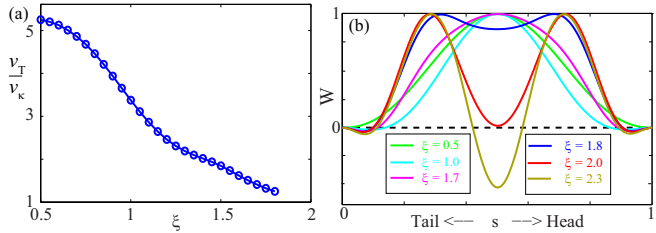}
  \caption{The speed of the internal torque wave normalized by the speed of the curvature wave as a function of the wave number (a) and the energy per cycle required as a function of the body position for different wave numbers (b).} \label{fig:power}
  	\end{center}
\end{figure}

The speed of the torque wave normalized to the speed of the curvature decreases from 5.2 to 1.3 as the wave number increases from 0.5 to 1.8 (Fig.~\ref{fig:power}a). This result is consistent with the results of previous studies, namely, the muscle activation is nearly synchronized for short wavelengths~\cite{wardle1995tuning}. The fit of a single traveling wave is poor when $\xi>1.8$; therefore, the wave speed is not defined and shown in Fig.~\ref{fig:power}a. 

The energy output per cycle required to overcome the external force at each point on the body is computed by integrating the power over a cycle, i.e., $W=\int_0^1 T \dot{\kappa}\mathrm{d}t$. The decrease in the phase difference between $T$ and $\dot{\kappa}$ and the decrease in the amplitude of $T$ at the middle as $\xi$ increases result in a more uniform distribution of the energy output over the body (see the blue line in ~\ref{fig:power}b). When $\xi=2$, the instantaneous power and energy output of the middle segment are zero. As $\xi$ further increases to 2.3, the power of the middle segment becomes negative, which means that the energy generated by other parts of the body is transferred to this part.

\begin{figure}
	\begin{center}
    \includegraphics[width=0.5\textwidth]{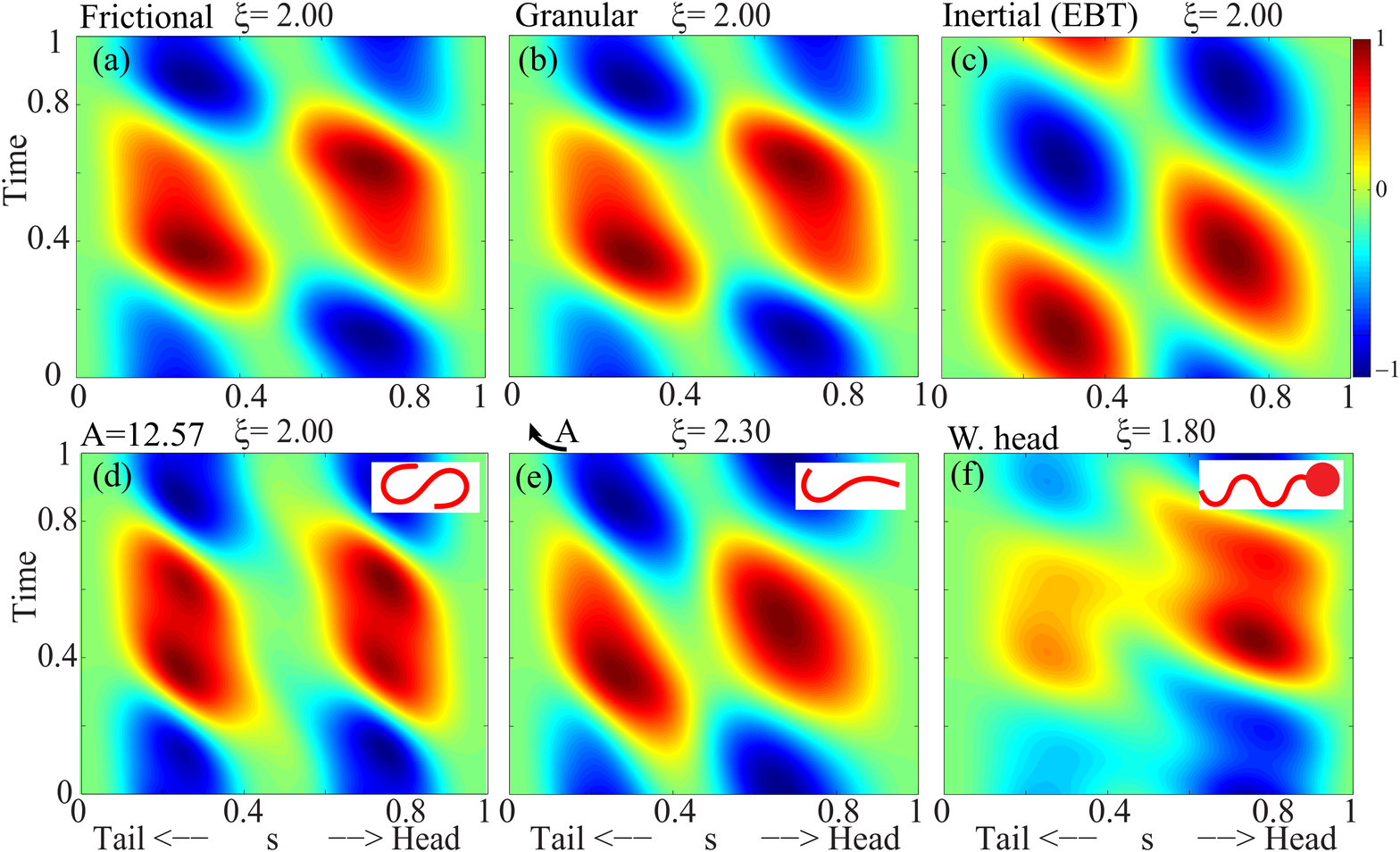}
  \caption{Torque pattern variation for different force laws, kinematics and geometry.  Torque pattern with (a) frictional, (b) granular (b), and (c) inertial  force laws. $A=7.54$ in (a-c).(d) Large amplitude $A=12.57$. (e) Increasing amplitude toward the tail. $A=7.54(1-s)$. (f) With a large head. Insets in (d-f) are schematic diagrams of the corresponding models. The head is not drawn to scale. See Supplementary Videos S3-8 for the respective torque patterns as a function of $\xi$ for (a-f).}\label{fig:var}
  	\end{center}
\end{figure}

A few variation are tested to evaluate the influences of the external forces and kinematics on the torque pattern and transitions. First, two additional types of resistive force laws obtained in previous experiments are considered: forces for granular media are described by $F_\perp=C_n \sin[\arctan(\gamma \sin(\phi))]$ and $F_\parallel=[C_f \cos(\phi) + C_l(1-\sin(\phi))]$, where $C_n=5.57$, $C_l=-1.74$, $C_f=2.30$, $\gamma=1.93$, and $\phi=\arctan(v_\perp/v_\parallel)$~\cite{maladen2011mechanical}. These force laws are empirical fitting functions for an aluminum cylinder dragged with different orientations in 3mm glass beads. For anisotropic frictional forces, $F_\perp=\mu_t v_\perp/|\mathbf{v}|$ and $F_\parallel=[\mu_f H(v_\parallel)+\mu_b(1-H(v_\parallel))]v_\parallel/|\mathbf{v}|$, where $\mu_f=0.3,\mu_b=1.3\mu_f$ and $\mu_t=1.8\mu_f$ are the friction coefficients in the forward, backward, and normal directions, respectively~\cite{hu2009mechanics}. $H(x)$ is the Heaviside step function. Since we focus on the torque pattern and normalize the torque by its maximum value, the absolute magnitudes of the forces are irrelevant here. These force laws and coefficients are obtained by measuring the frictional forces while unconscious snakes slide on clothes at different orientations. The torque pattern is qualitatively the same when the force laws are replaced by those for granular and frictional environments, and only subtle differences are observed (Fig.~\ref{fig:var}a\,\&\,b). We also compute the force distribution using Lighthill's elongated body theory (EBT) where only the lateral inertial forces from the fluid are considered (the derivation is provided in the Supplementary Information)~\cite{lighthill1960note}. The inertial forces considered in the EBT give a torque pattern that is similar to that from the RFT, albeit with a phase shift of $+\pi/2$ (1/4 period) (Fig.~\ref{fig:var}d). 

To study the effect of undulation amplitude, we increase the amplitude to $A=12.57$, at which the segments nearly overlap. Surprisingly, we find that the torque pattern is insensitive to amplitude (see Fig.~\ref{fig:var}d and Supplementary Video S9).  When the amplitude increases linearly toward the tail, similar transitions occur, albeit at a greater wave number (Fig.~\ref{fig:var}e). Since interactions between segments through the medium are neglected, the results from the amplitude variations only include the geometric effects. An example case with head drag is also examined (Fig.~\ref{fig:var}f). Based on an experiment on bull spermatozoa, the head is approximated as a sphere with isotropic drag, and the drag coefficient is chosen such that the head drag is 58\% of the body drag when the body is straight and moving perpendicular to its axis~\cite{friedrich2010high}, i.e. $\mathbf{F}_h=0.58C_\perp \mathbf{v}(1)$. The resulting torque is significantly larger for the anterior half of the body and the transition of torque to the two-wave pattern occurs at a smaller wave number. 

\begin{figure*}
	\begin{center}
    \includegraphics[width=0.9\textwidth]{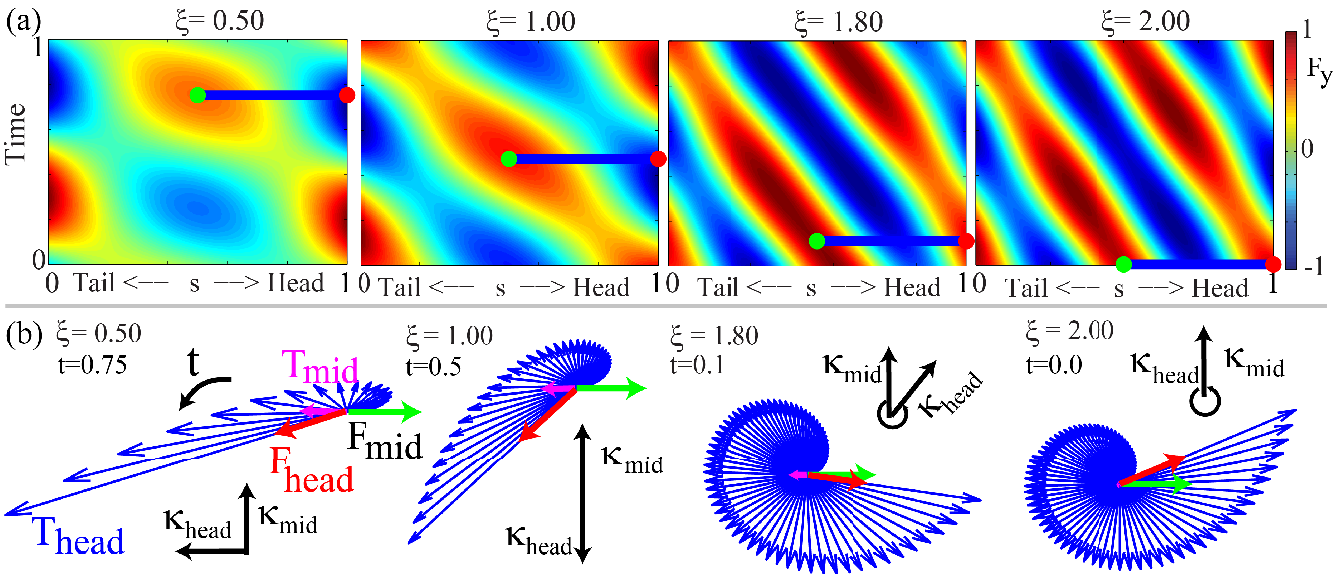}
  \caption{Composition of the torque at the middle and phasor diagrams. (a) The lateral force $F_y$ distribution as a function of body position ($s\approx x$) and time. (b) Phasor diagrams.  The blue arrows represent the contribution of the force on the anterior part of the body to the torque at the middle of the body when the force at the middle is maximum. The integration regions are marked by the thick blue lines in (a). $T^e_\mathrm{mid}$ and  $T^e_\mathrm{head}$ represent the torque at the middle and the head, respectively, and the corresponding forces are marked by green and red dots in (a). The black arrows represent the curvature phasors. The lengths of the force, torque, and curvature phasors are drawn to reflect only their relative magnitudes for the same value of $\xi$.}\label{fig:forcedis}
  	\end{center}
\end{figure*}

To elucidate the mechanism underlying the transition of the torque, we analyze the phases of the torque at the middle point of the body and at a point infinitely close to the head as examples (Fig.~\ref{fig:forcedis}). In a previous study~\cite{ding2013emergence}, the torque at the middle point was roughly decomposed into three parts to explain the neuromechanical phase lag. Here we use a more quantitative tool--phasor diagram--to visualize and analyze the relationships among the phases of curvature, force and torque.

Further simplification is needed prior to the analysis. Since the torque pattern is not sensitive to amplitude, a small amplitude ($A=0.6$) is used such that the locomotor is nearly a straight line on the $x$ axis undulating in place. In this case, only the lateral displacement ($y$) and lateral forces ($F_y \approx F_\perp $) need to be considered and the longitudinal forces ($\approx F_\parallel$) are negligible. Then the equation for computing the torque at position $s$ can be simplified as $T(s)=-\int_s^1 (l-s) F_y \mathrm{d}l$. Prior to the analysis, we also note that the spatio-temporal patterns of the lateral force are affected by the requirements of force and torque balances. For $\xi=0.5$, a wave number less than 1, the phase difference between the lateral forces at the head and at the middle point is greater than $\pi/2$, which is the phase difference of the curvatures at these two points (Fig.~\ref{fig:forcedis}). This result can be understood by considering the balance of the lateral forces and the torque: zero total lateral force requires both negative forces and positive forces to be present at any time; the zero torque condition further requires that the negative forces be distributed on both sides when the force in the middle is positive (a similar argument was made by Gray \cite{gray1946mechanism}). Nonetheless, the phase differences of the forces between the middle point and the end points increase with increasing wave number.

In phasor diagrams, a variable is represented by a phasor (vector), whose projection on the horizontal axis is the instantaneous value of the variable and the rotation of the phasor corresponds to the time evolution. Since $T$ and $T^e$ only have a sign difference, we examine the phasors of the external torque $T^e$ and first focus on the phasor of $T^e$ at the middle point of the body ($T^e_\mathrm{mid}=T^e(s=0.5)$ in Fig.~\ref{fig:forcedis}b). The force contribution to $T^e$ at the middle point (i.e. $(l-0.5)F_y$ for $l>0.5$) is discretized and visualized using phasors in Figure \ref{fig:forcedis}b. The integrative nature of the torque at the middle point makes the phase of $T^e$ between the phases of the forces. Interestingly, the torque $T^e$ at the middle point is precisely either out of phase (for $\xi<2$) or in phase (for $2<\xi<2.3$) with the force at the middle point. This alignment can be understood by considering the symmetry and torque balance when the force at the middle is zero (this case is similar to the one shown in Fig.~\ref{fig:model} with the fore-aft forces ignored): the lateral displacement is symmetric about the middle point while the lateral velocity and force distributions are antisymmetric. The antisymmetric force distributions generate torque about the middle point with the same sign but the total torque on the body must be zero. Therefore, the torque from each half of the body (i.e. $T^e(0.5)$) at this time instant must be zero. When $\xi$ approaches 2, one full wavelength appears on each side of the body, the torque contributions cancel out, and $T^e_\mathrm{mid}$ becomes zero. As $\xi$ continues to increase, the $T^e_\mathrm{mid}$ becomes in phase with the local force $F_\mathrm{mid}$. This situation corresponds to a breakdown in the torque pattern of a single traveling wave and a reversal of the local torque wave at the middle. 

To understand the speed variation of the torque wave (Fig.~\ref{fig:power}a), we compare the phase differences of the torque, force and curvature at the middle point and at the head ($s=1$). At a point infinitely close to the head, the torque $T^e_\mathrm{head}$  is simply in phase with the local force at the head ($F_\mathrm{head}$). Therefore, the phase difference between the torques at the middle point and at the head (the angle between $T^e_\mathrm{mid}$ and $T^e_\mathrm{head}$ in Fig.~\ref{fig:forcedis}b) is considerably smaller than the phase difference between the forces (the angle between $F_\mathrm{mid}$ and $F_\mathrm{head}$ in Fig.~\ref{fig:forcedis}b) and the phase difference between curvatures (the angle between $\kappa_\mathrm{mid}$ and $\kappa_\mathrm{head}$ in Fig.~\ref{fig:forcedis}b). As $\xi$ increases from 0.5 to 1.8, the phase of $T^e_\mathrm{mid}$ remains the same, but $T^e_\mathrm{head}$ increases at the same rate of $F_\mathrm{head}$. Therefore, the phase difference of torque increases from a smaller number (approximately $0.25\pi$ for $\xi=0.5$) to nearly $\pi$. Since such an increase is greater in proportion compared to the increase in curvature, which is from $0.5\pi$ to $1.8\pi$, the speed of the torque wave relative to the curvature wave decreases.  

As shown in the above analysis, the torque pattern is primarily determined by the phase distribution of the forces modulated by distance. This picture can also help explain the observed torque variations (Fig.~\ref{fig:var}). When the force laws are changed to granular or frictional ones, the phase distributions of the forces on the body are similar; therefore, similar torque patterns are observed. The phase of the reactive force is proportional to the time derivative of the velocity and is hence ahead of the phase of the resistive force by $\pi/2$; therefore, the phase of the torque is shifted by the same amount. When the curvature amplitude increases toward the tail, the motion and forces on the head are relatively small; thus, the effective phase range from the head to tail is smaller than the nominal one indicated by $\xi$, and the two-wave transition is delayed (greater $\xi$). For the case with a head, because the head force is nearly out of phase with the total torque from other parts when the two-wave transition occurs (see the red and magenta arrows in Fig.~\ref{fig:forcedis}b, $\xi=1.8$), the enhancement of the head drag causes the cancellation and reversal of the torque to occur earlier (smaller $\xi$).

\section{Discussion}
\begin{figure}
	\begin{center}
    \includegraphics[width=0.5\textwidth]{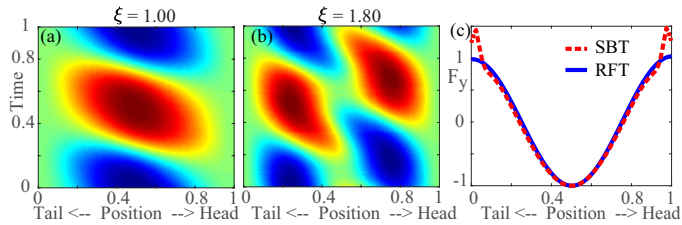}
  \caption{Torque pattern computed using slender body theory for viscous fluids. (a) \& (b) are the torque patterns for different wave numbers. See Supplementary Video S10. $A=7.54$. (c) Comparison of the lateral force along the body using resistive force theory and slender body theory. $t=0$, and $A=0.6$.}\label{fig:SBT}
  	\end{center}
\end{figure}
In the resistive force theory for viscous fluids, the assumption that the force on one segment is independent of the movement of other segments might introduce significant errors in the forces~\cite{rodenborn2013propulsion}. Such error can be alleviated by using slender body theory~\cite{lighthill1976flagellar}. In slender body theory, the body of the swimmer is assumed to be slender, and the ratio between the radius of the body and body length $a/L$ is much smaller than 1. Singularity solutions of point forces and dipoles are arranged along the body centerline ,and the velocity at a point is computed as the superposition of the singularity solutions to include the effect of the interaction between segments (see \cite{rodenborn2013propulsion} for the details of the explanation and implementation). Here, we use a biologically relevant body shape $1/L=1/30$~\cite{moore2013wormsizer} and the same kinematic parameters in RFT. We found that the transition of the torque pattern from slender body theory is qualitatively the same as those from resistive force theory but the transition to the two-wave pattern occurs at smaller $\xi$ ($\approx$1.8) (Fig.~\ref{fig:SBT}a\,\&\,b). The wave speed ratio from SBT is also slightly smaller than the result from RFT (e.g., 3.0 vs. 3.3 at $\xi=1$). Examination of the force distribution for a small amplitude reveals one mechanism for the early transition: forces at the head and tail are larger because at the ends, the segments experience greater drag force as less segments are nearby to ``help'' induce the flow. Similar to the case with a head (Fig.~\ref{fig:var}f), the head and tail forces are nearly out of phase with the total torque from other parts and the enhancement of the drags and the ends causes the cancellation and reversal of the torque to occur earlier.

\begin{figure}
	\begin{center}
    \includegraphics[width=0.5\textwidth]{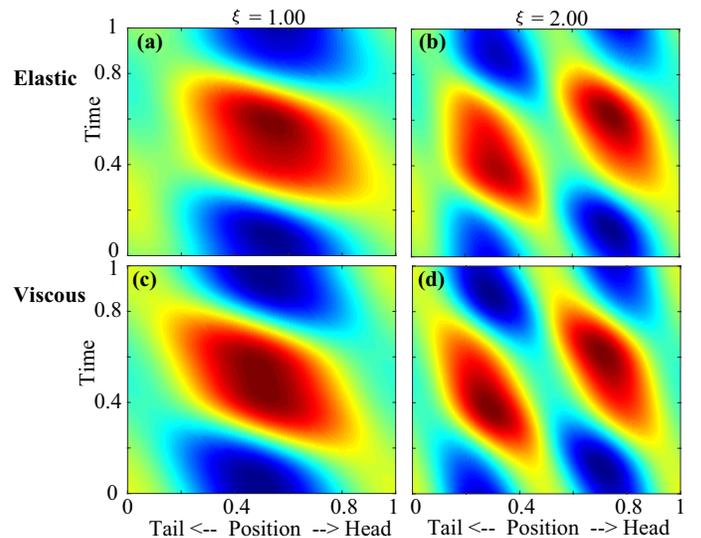}
  \caption{Torque pattern variation when the elastic or viscous body forces are included. The four subfigures represent combination of two kinds of forces and two the wave numbers. See Supplementary Videos S11-12
for the respective torque patterns as a function of $\xi$ for (a,b) and (c,d).}\label{fig:elast}
  	\end{center}
\end{figure}
The resistance to the bending from the body can also contribute to the torque, therefore, we further discuss the effects of elastic and viscous forces in the body in a general way. We assume that the elastic force requires an additional torque $T_s=C_e\kappa$ and that the viscous force requires an additional torque $T_v=C_v\dot{\kappa}$. Since when these torque dominates, the torque pattern just coincides with the curvature pattern, a traveling wave, we consider the case in which these torques are significant but smaller compared to the torque from external forces. Therefore, the coefficients $C_e$ and $C_v$ are chosen such that maximal values of these torques are 20\% of the maximal values of the torque from external forces, i.e. $T=T^e/\max(T^e)+0.2 \dot{\kappa}/\max(\dot{\kappa})$ and $T=T^e/\max(T^e)+0.2 \kappa/\max(\kappa)$. For $\xi<1.8$, the inclusion of elastic force causes the wave speed of the torque in the middle part to decrease (Fig.~\ref{fig:elast}a). For example, $v_T/v_\kappa= 2.9$ for $\xi = 1$. The inclusion of viscous force causes the wave speed of the torque in the middle part to decrease. For example, $v_T/v_\kappa= 1.8$ for $\xi = 1$. 

In this study, we also adopted a highly simplified locomotion gait, but organisms adopt gaits different from a single-mode sinusoidal curvature wave during turning and other maneuvers~\cite{padmanabhan2012locomotion,saggiorato2017human}. The torque pattern and neural control required for these maneuvers may be quite different and warrant further study.

As shown in our variation study and previous studies, inertia, body elasticity, interactions between body parts, and complex body geometry may all affect the torque and muscle activation patterns. Therefore, the predicted torque from our simple model probably cannot match the muscle activation of a particular organism in detail. However, the torque predicted by our model is certainly an important part of the total torque that needs to be overcome by many organisms. 
  
Our results predict that muscle activation is no longer a traveling wave when the dominant forces that the animal must overcome are external resistive forces and the wave number is greater than two (e.g., the snake in \cite{sharpe2015locomotor}). However, to our knowledge, muscle activation and neural control in animals with wave numbers greater than 2 have not been studied. From another perspective, our results predict that muscle activation of a traveling wave cannot produce a uniform bending wave for more than two wavelengths if external forces significantly contribute to the torque. For robotic systems, our results show that the distributions of torque magnitude and energy output along the body can be adjusted by varying the wave number; this information may guide the design of driving systems and the use of passive materials.

In summary, our study provides a general picture of the torque pattern from resistive forces in undulatory locomotion, including new and complex patterns that have not previously been observed. By introducing the phasor diagram for undulatory locomotion, we show that the torque pattern can be understood from the integration of distance-modulated force phasors and that the rapid transitions occurring near integer numbers are the result of the cancellation of the force phasors. The phasor diagram method may be a useful tool to further investigate the interplay between torque, passive body forces, body shape, and external forces in undulatory locomotion.

\section{Acknowledgments}
Funding for Y.D. and T.Y.M. was provided by NSFC grant No. 11672029, NSAF-NSFC grant No. U1530401, and the Recruitment Program of Global Young Experts.

\bibliographystyle{apsrev4-1}

\end{document}